# GRAPH STRUCTURAL COMPLEXITY

A.A. Snarskii[1,2]

[1]*The National Technical University of Ukraine "Igor Sikorsky Kyiv Polytechnic Institute", Dep. of General Physics, Kyiv, Ukraine*

[2]*Institute for Information Recording, NAS of Ukraine, Mykoly Shpaka Street 2, 03113 Kiev, Ukraine*

Abstract.

Introduced the quantitative measure of the structural complexity of the graph (complex network, etc.) based on a procedure similar to the renormalization process, considering the difference between actual and averaged graph structures on different scales. The proposed concept of the graph structural complexity corresponds to qualitative comprehension of the complexity. The proposed measure can be obtained for the weighted graphs also.

The structural complexities for various graph types were found – the deterministic infinite and finite size graphs, artificial graphs of different natures including percolation structures, and the time series of cardiac rhythms mapped to complex networks using the parametric visibility graph algorithm. The latter reaches a maximum near the formation of a giant component in the graph or at the percolation threshold for 2D and 3D square lattices when a giant cluster having a fractal structure has emerged. Therefore, the graph structural complexity allows us to detect and study the processes similar to a second-order phase transition in complex networks.

A new node centrality index, characterizing the structural complexity of a certain node within the graph structure is introduced also, it can serve as a good auxiliary or generalization to the local clustering coefficient. Such an index provides another new ranking manner for the graph nodes.

Being an easily computable measure, the graph structural complexity might help to reveal different features of complex systems and processes of the real world.

## 1. INTRODUCTION

The term complexity stands for many different concepts. In [1-4] one can find a vast list of already established types of complexities. They can be divided into groups: difficulty of description, difficulty of creation, computational complexity, etc. The complexity analysis of many natural phenomena reveals internal processes occurring in the investigated system often impenetrable by commonly established descriptive methods and algorithms. This justifies an increasing trend towards the use of complexity analysis in quantifying different real-world data, for example, the neural activity measured by electroencephalography (EEG) signals [5], etc.

Recently, a new kind of complexity was introduced – the Multiscale structural complexity (MSC) [6], which deals with the complexity of various types of colorings or patterns in multidimensional lattice space. These include, for example, two-dimensional images of various structures, from images of nature to stripe domains in ferromagnetic thin films, pearlitic structure in rail steel, stripes on a beach in a tide zone, and much more. In other words, the MSC deals with the weights of the nodes of a regular spatial lattice, for instance, a two-dimensional square lattice for ordinary images. It has been shown [6] that the MSC reveals important information about the dynamics of nonequilibrium structures, the phase transition phenomena, and much more.

In the case of a two-dimensional black-and-white image of size $L \times L$, $L = 2^n$ represented by a function $f(x,y)$ with an average value $\langle f(x,y) \rangle$, the algorithm [6] consists of $n$ steps, and at each step a value called deviation to be found. For the first step, i.e. for the whole image, it is the variance $D_1 = D^1(L) = \left\langle \left( f(x,y) - \langle f(x,y) \rangle \right)^2 \right\rangle$. At the second step the image is divided into four equal parts of size $L/2 \times L/2$ and the variance of each part is calculated $D^i(L/2), \quad i = 1..4$ and then the deviation is found as the sum of the variances $D_2 = \sum D^i \left( L/2 \right)$. The splitting into smaller parts $L/4, L/8, ...$ down to the minimum size, i.e. pixel size and applying to each step a similar procedure for deviation we find the series $D_1, D_2, ... D_n$. The MSC according to [6] is the sum of the deviations $D = \sum D_i$. Thus, the MSC value characterizes the difference between the image pixel intensities and averaged intensity data at different scales.

Note that there is a procedure close to the described MSC algorithm for analyzing a one-dimensional “image”, i.e. a data series or time series – the detrended fluctuation analysis [7, 8]. In this case, the series $f_i$ is divided into segments of the length $k$ and the function $y(k) = \sum_{i=1}^{k} \left( f_i - \langle f_i \rangle \right)$, where $\langle f_i \rangle = \frac{1}{k} \sum_{i=1}^{k} f_i$ is constructed and then its linear approximation $y_n(k) = y_{0n} + C_n k$ is built (in the MSC algorithm [6] the average is calculated only). Finally, the root-mean-square fluctuation $F(n)$ is found

$$F(n) = \sqrt{\frac{1}{N} \sum_{k=1}^{N} \left[ y(k) - y_n(k) \right]^2} \sim n^{\alpha}, \qquad (1)$$

where $\alpha$ - the index of the interest. For example, for a time series of the human heart rate intervals $\alpha = 1$ and for persons with cognitive heart failure $\alpha = 1.3$ [7, 8].

In this work, we introduce the Graph Structural Complexity (GSC, GSComplexity) using an approach similar to [6]. The GSC algorithm consists of sequential calculating the Node Structural Complexity (NodeSC) for each node of the graph followed by their averaging over all graph (complex network, regular lattice, etc.) nodes. To calculate the NodeSC, we propose the algorithm for dividing a graph into parts/subgraphs, it is based on the breadth-first search algorithm [9], where for an arbitrary node at the $k$-th step of the algorithm, all the reachable graph nodes and the edges connecting them make up a subgraph.

Let us denote the number of nodes of the subgraph as $s_k$ and the number of edges – $b_k$, so the density of the subgraph formed at each step of the algorithm will be $p_k = b_k / m_k$, where $m_k = s_k(s_k - 1)/2$ is the number of edges in the complete graph with $s_k$ nodes. Also, we assign for the edges of the subgraph a unity weight and the weight $p_k$ (the density) to the edges of the corresponding complete graph. Now for every $k$-th step of the algorithm it will be possible to calculate the deviation of the weights of the subgraph and the complete graph as follows

$$\delta_k = \frac{1}{m_k}\left[\sum(1-p_k)^2 + \sum(0-p_k)^2\right], \qquad (2)$$

where the first term deals with $b_k$ edges presented both in the subgraph and the complete graph, and the second one – with the complementary to the complete graph $m_k - b_k$ edges (i.e. “missing” in subgraph edges, considering their weight is equal to zero).

The NodeSC is the sum of deviations $\delta_k$ calculated for all possible steps of the breadth-first search algorithm, and the GSC is the average of the NodeSC over all the nodes of the graph.

The subgraph at the $k$-th step of the breadth-first search algorithm can be considered as the analog to the part of the image from [6] or a segment from DFA [3,4], and the subgraph density as the analog to the average of the image part in [6].

Further, depending on the context, we will use both the terms graph (mainly when we are talking about a deterministic, relatively simple structure) and the term network (mainly when we are talking about a large graph formed with elements of randomness) or lattice, when we are supposed to construct a regular graph or lattice-based graph.

The work consists of 6 chapters. This first chapter is an introduction. In the second chapter, the algorithm of the graph structural complexity will be described in detail. In the third chapter, the complexities of several deterministic graphs will be calculated. In the fourth – the complexity of random artificial graphs. The fifth chapter will be devoted to random graphs generated by a time series of cardiac intervals mapped into a network by the parametric viability graph algorithm [10, 11]. In the last part, we will discuss the results and perspectives of the GSC algorithm. Addendums A and B describe in detail results for certain types of graphs.

## 2. THE GSC ALGORITHM

We will consider an arbitrary graph $G=(\Sigma,\mathrm{B})$ where $\Sigma$ is a set of nodes having a cardinality $Q$, $\mathrm{B}$ is a set of edges. The graph structural complexity – $C(G)$ is the average of the nodes structural complexity $C^i(G)$ of all graph nodes

$$C(G)=\frac{1}{Q}\sum_{i=1}^{Q}C^i(G). \qquad (3)$$

Consider an arbitrary graph node $i\in[1..Q]$ of the graph $G$ and use a breadth-first search algorithm starting at the $i$-th node to construct at each step of the algorithm a subgraph $g_k^i(\sigma_k^i,\beta_k^i)$, $k=1...n^i$, where $\sigma_k^i$ is the set of nodes of the graph, reachable in no more than $k$ steps from the initial $i$-th node, total $s_k^i$ nodes, and $\beta_k^i$ is the set of all the graph edges connecting $\sigma_k^i$ nodes, in the total $b_k^i$ edges, $n^i$ denote the number of steps of the breadth-first search algorithm starting from the initial $i$-th node.

A set of subgraphs $g_k^i$ constructed for each $i$-th node at each $k$-th step will be used as an analog of the network partitioning during the renormalization procedure.

In each constructed subgraph $g_k^i$ we assign to all edges, total $b_k^i$ edges, the weight equal to 1, and to the remaining connections representing the complement $g_k^i$ to the complete graph $\bar{g}_k^i$, total $\bar{b}_k^i$ edges, a weight equal to 0, the number of edges of the complete graph $m_k^i=b_k^i+\bar{b}_k^i=s_k^i\left(s_k^i-1\right)/2$, the weight of the edge of the complete graph is taken equal to the density of the subgraph $g_k^i$, i.e.

$$p_k^i=b_k^i/m_k^i. \qquad (4)$$

Changes in the weights $\Delta_k^i$ and $\bar{\Delta}_k^i$ for each edge from $g_k^i$ and $\bar{g}_k^i$, will have different signs

$$\Delta_k^i = b_k^i - p_k^i \equiv 1 - \frac{b_k{}^i}{m_k{}^i},\ \bar{\Delta}_k^i = \bar{b}_k^i - p_k^i \equiv 0 - \frac{\bar{b}_k^i}{m_k^i}. \tag{5}$$

Thus, the variance of the weights of the subgraph edges $g_k^i$ is equal to

$$\delta_k^i = \frac{1}{m_k^i}\left( b_k^i \left(\Delta_k^i\right)^2 + \bar{b}_k^i \left(\bar{\Delta}_k^i\right)^2 \right) = p_k^i \left(1 - p_k^i\right). \tag{6}$$

Let's define the NodeSC of the $i$-th node as

$$C^i = \sum_{k=1}^{n_i} \delta_k^i. \tag{7}$$

Thus, the GSC of the entire graph is as follows

$$C(G) = \frac{1}{Q}\sum_{i=1}^{Q}\sum_{k=1}^{m_i} \delta_k^i = \frac{1}{Q}\sum_{i=1}^{Q}\sum_{k=1}^{m_i} p_i^k \left(1 - p_i^k\right). \tag{8}$$

According to (5), a complete graph and a graph consisting only of nodes (without edges) will have zero GSComplexity, and the GSComplexity of a graph $G_{G_1 \cup G_2}$ consisting of two subgraphs without joining edges is equal to

$$C\left(G_{G_1 \cup G_2}\right) = \frac{q_1 C(G_1) + q_2 C(G_2)}{q_1 + q_2}, \tag{9}$$

where $C(G_1)$ and $C(G_2)$ are the GSC of each subgraph, $q_1$ and $q_2$ are the number of subgraph's $G_1$ and $G_2$ nodes.

Let us consider the NodeSC calculation for node #1 using the example graph $G_{Ex}$ shown in Fig. 1. The figure shows the nodes and edges of the first three steps of the GSC algorithm (there are only three steps for node #1 in the graph $G_{Ex}$ ). Hereinafter, the figures will use to show the first three steps of the GSC algorithm starting at the marked node #1 and to use the same line types and colors (oline) for the edges involved.

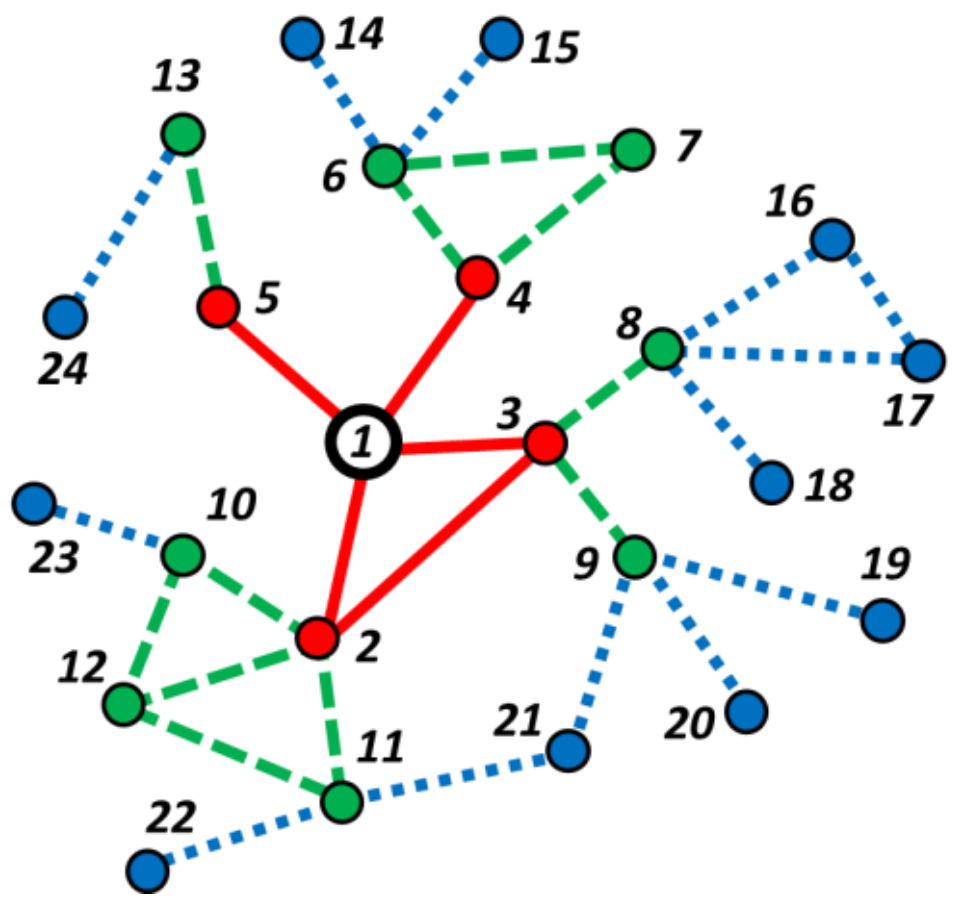

Fig.1. The graph $G_{Ex}$ to explain the GSComplexity algorithm for node #1 with marked edges for the first three steps of the algorithm. (red, green and blue online).

In the first step, for node #1 $\sigma_1^1 = \{1,2,3,4,5\}$ and the edges set connecting them is $\beta_1^1 = \{(1-2),(1-3),(1-4),(1-5)\ ,(3-2)\}$ (red solid line in Fig. 1). Thus, in the first step the subgraph consists of five nodes and five edges $s_1^1 = 5$, $b_1^1 = 5$. The number of edges in the complete subgraph is $m_1^1 = s_1^1\left(s_1^1 - 1\right)/2 = 10$, its density (4) and variance (6) are $p_1^1 = b_1^1 / m_1^1 = 1/2$, $\delta_1^1 = p_1^1\left(1 - p_1^1\right) = 1/4$, accordingly. At the second step, the set of nodes of the subgraph is $\sigma_2^1 = \{1,2,...,13\}$ and $s_2^1 = 13$, and the set of edges is $\beta_2^1 = \beta_1^1 + \{(2-10),(2-11),(2-12),(11-12),(10-12),(3-8),(3-9),(4-6),(4-7).(6-7),(5-13)\}$

(red solid and green dashed lines) and $b_2^1 = 16$. Thus, the density is $p_2^1 = b_2^1 / m_2^1 = 16/78$ and the variance are equal to $\delta_2^1 = p_2^1\left(1 - p_2^1\right) = 1662/78^2 \approx 0.163$. The edges added at the third step are blue dotted lines in Fig.1. In total, it takes 3 steps to calculate the NodeSC of node #1, $C^1(G_{Ex}) \approx 0,601$, and the GSComplexity of the entire graph GSC is $C(G_{Ex}) \approx 0.885$

Table 1. Node structural complexities for the graph $G_{Ex}$ in Fig. 1

| Node # | NodeSC | Node # | NodeSC | Node # | NodeSC |
|---|---|---|---|---|---|
| 1 | 0.601 | 9 | 0.876 | 17 | 0.904 |
| 2 | 0.708 | 10 | 0.911 | 18 | 0.904 |
| 3 | 0.685 | 11 | 0.883 | 19 | 0.876 |
| 4 | 0.753 | 12 | 0.912 | 20 | 0.876 |
| 5 | 0.784 | 13 | 1.034 | 21 | 1.036 |
| 6 | 1.021 | 14 | 1.021 | 22 | 0.912 |
| 7 | 0.771 | 15 | 1.021 | 23 | 0.911 |
| 8 | 0.904 | 16 | 0.904 | 24 | 1.034 |

The GSC algorithm can be generalized to the case of weighted graphs $G = (\Sigma, \mathrm{B}, \Omega)$, where in the definition of the graph a set of edges weights $\Omega$ is added, the set of the edges weights $g_k^i$ at the $k$-th step of the algorithm is denoted as $\omega_k^i$, and the weight of the edge $j$ of this set – $w_k^i(j)$. When calculating the GSComplexity of such a graph, the relations (4)-(8) have become somewhat more complicated. The density of the subgraph $g_k^i$ is found by analogy with (4)

$$p_k^i = \frac{1}{m_k^i} \sum_{j=1}^{b_k^i} w_k^i(j). \qquad (10)$$

For each edge from the set $\beta_k^i$, it is possible to calculate the weight differences – $\Delta_k^i(j)$, for all missing edges the value $\overline{\Delta}_k^i$ is set to (5).

$$\Delta_k^i(j) = w_k^i(j) - p_k^i,\ j = 1..b_k^i,\ \ \bar{\Delta}_k^i = 0 - p_k^i. \qquad (11)$$

Accordingly, the variance of the edge weights of the original graph $g_k^i$ is equal to a sum similar to (3)

$$\delta_k^i = \frac{1}{m_k^i}\left(\sum_{j=1}^{b_k^i}\left(\Delta_k^i(j)\right)^2 + \bar{b}_k^i\left(\bar{\Delta}_k^i\right)^2\right). \qquad (12)$$

The relationship for the NodeSC (7) remains unchanged, and the relationship (8) becomes

$$C(G) = \frac{1}{Q}\sum_{i=1}^{Q}\sum_{k=1}^{n_i}\delta_k^i = \frac{1}{Q}\sum_{i=1}^{Q}\sum_{k=1}^{n_i}\left(\frac{b_k^i}{m_k^i}\left\langle\left(w_k^i(j)\right)^2\right\rangle - \left(\frac{b_k^i}{m_k^i}\right)^2\left\langle w_k^i(j)\right\rangle^2\right), \qquad (12)$$

where $\langle\ \rangle$ denotes the arithmetic mean.

Thus, the GSC algorithm also allows us to find the GSComplexity of the weighted graphs.

If the edges in the graph have equal weight, then the GSC of the complete graph will be zero, but if the edges have different weights, then even the complete graph will have non-zero GSComplexity.

# 3. COMPLEXITY OF DETERMINISTIC GRAPHS

## 3.1 Graphs composed of complete graphs

The graphs composed of the complete subgraphs that do not have common nodes have the GSC equal to zero, it directly follows from (7) and (8). There are also some other complete graph compositions allowing the GSC value manual calculations.

Let's consider a graph $G_{G_1+G_2}(Q)$ consisting of two complete subgraphs $G_1(q_1)$ and $G_2(q_2)$ with one common node, see Fig. 2a, in this case the total number of the graph nodes is $Q = q_1 + q_2 - 1$.

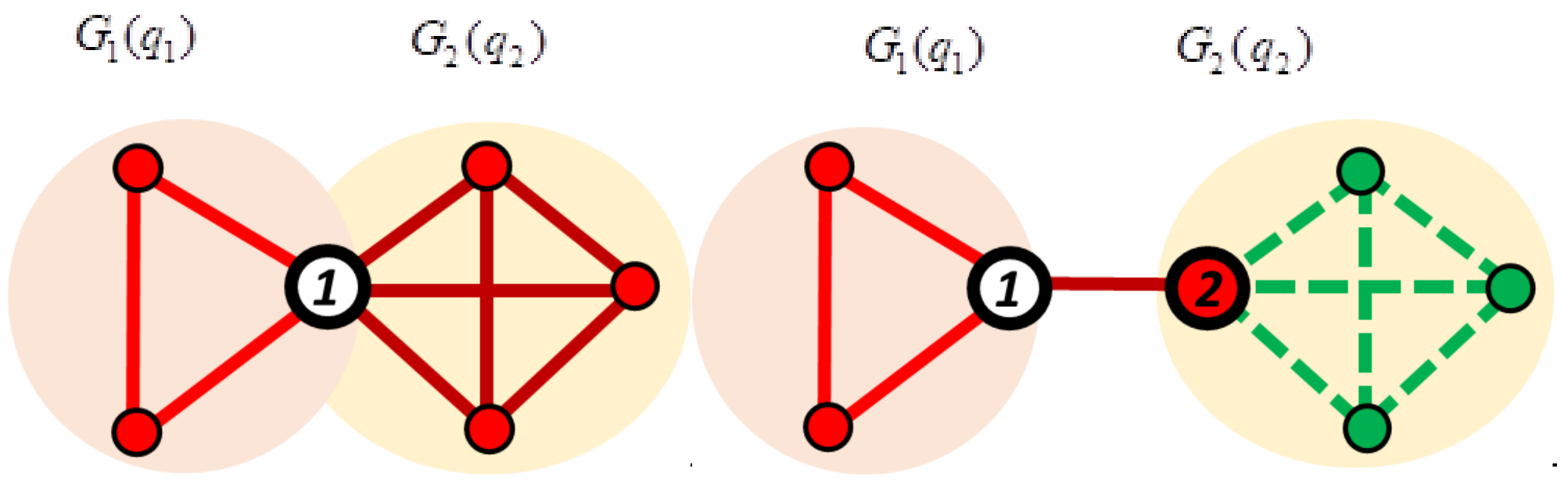

a b

Fig.2. Graphs composed of complete subgraphs, steps of the GSC algorithm for node #1 are marked, and the complete subgraphs are highlighted. a – two complete subgraphs having one common node, b – two complete graphs connected by a single edge (color online).

The graph $G_{G_1+G_2}(Q)$ Fig. 2a has two types of nodes: type 1 – the nodes belonging to only one of the subgraphs and type 2 – node #1 marked in Fig. 2a, belonging simultaneously to both subgraphs. For node #1 (type 2), the number of steps of the GSC algorithm is 1, and therefore its NodeSC is

$$C^1(G_{G_1+G_2}) = \delta_1^1 = p_1^1(1-p_1^1), \quad p_1^1 = \frac{q_1(q_1-1)+q_2(q_2-1)}{Q(Q-1)}. \tag{13}$$

For the nodes of type 1, the total number of the GSC algorithm steps is 2, $\delta_1^i = 0, \delta_2^i = \delta_1^1, i = 2..Q$ and thus the NodeSC for them is

$$C\left(G_{G_1+G_2}\right) = 2\left(q_1-1\right)\left(q_2-1\right)\frac{q_1\left(q_1-1\right)+q_2\left(q_2-1\right)}{\left(q_1+q_2-1\right)^2\left(q_1+q_2-2\right)^2}, \tag{14}$$

the (14) is valid for $q_1+q_2>2$, for the smaller subgraph sizes the GSC value can be found directly using (4) - (8): $C\left(1,1\right)=0$, $C\left(1,q_2\right)=0$.

If the complete subgraphs have the same number of nodes, $q_1=q_2=q$ then the GSC is

$$C\left(G_{2G_1}\right) = q\frac{q-1}{\left(2q-1\right)^2} \tag{15}$$

and at large node numbers, the GSC tends to $C\left(q \gg 1\right) \to 1/4$.

It should be noted also, that in the case of $q_1 >> q_2$ the GSC (15) will tend to zero, i.e. to the GSComplexity of a complete graph. Thus, at least in this case, a small number of additional nodes/edges will not affect the GSComplexity of the whole graph of a large size.

The graph $G_{G_1+1+G_2}(Q)$ Fig. 2b consists of the two complete subgraphs $G_1(q_1)$ and $G_2(q_2)$ connected by a single additional edge. The total number of nodes in that graph is $Q=q_1+q_2$. In such a graph there are four types of nodes: type 1 and type 2 - nodes belonging to one of the complete subgraphs, but not belonging to the edge connecting the subgraphs (1-2), type 3 and type 4 – nodes belonging to the connecting edge, indicated in Fig. 2b node #1 and #2, the GSC of such a graph is

$$C\left(G_{G_1+1+G_2}\right)=\frac{2}{q_1+q_2}\left\{\frac{(q_1-1)\left[q_1(q_1-1)+2\right]}{q_1(q_1+1)^2}+\frac{(q_2-1)\left[q_2(q_2-1)+2\right]}{q_2(q_2+1)^2}+(q_1q_2-1)\frac{q_1(q_1-1)+q_2(q_2-1)}{(q_1+q_2)(q_1+q_2-1)^2}\right\} \quad (16)$$

Note that the (16) is symmetric relative to $q_1 \Leftrightarrow q_2$, i.e. $C\left(G_{G_1+1+G_2}\right)=C\left(G_{G_2+1+G_1}\right)$,.

Fig. 3 shows the GSC of the graph $G_{G_1+1+G_2}(Q)$ composed of the subgraphs of different cardinality. As the number of the subgraph nodes increases, two cases can be considered. In the case $q_1=q_2=q\to\infty$, the GSComplexity value (16) tends to $\lim_{q\to\infty}C\left(G_{G_1+1+G_1}\right)=1/4$, as (15) does for GSC of $G_{G_1+G_2}(Q)$. In the second case $q_1 \gg q_2$ and the GSC falls inversely proportional to $q_1$, namely $C\left(q_1\to\infty,q_2\right)\sim 2\left(1+q_2\right)\cdot q_1^{-1}$. Fig. 3 shows the GSC of $G_{G_1+1+G_2}(Q)$ behavior and demonstrates both described cases.

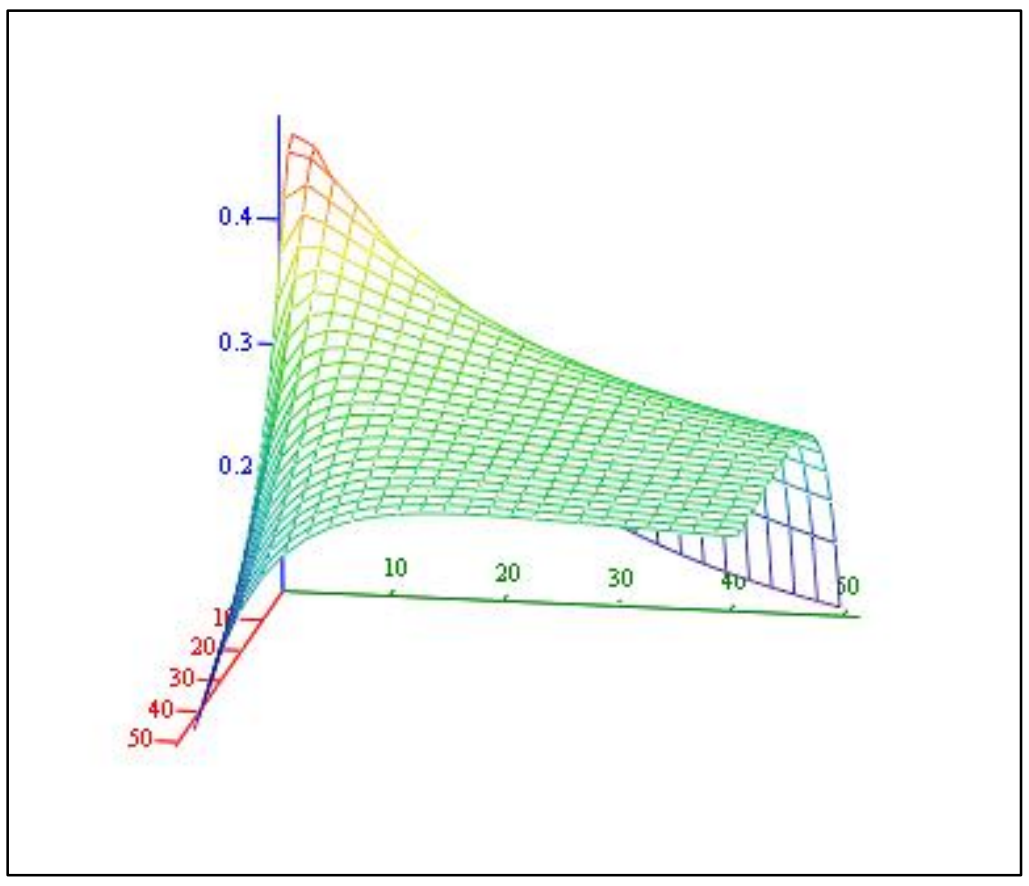

Fig.3. The GSComplexity of the graph $G_{G_1+1+G_2}(Q)$ composed of the complete subgraphs connected by a single edge.

Consider a complete graph without one edge $G_{q-1}(Q)$, the total number of nodes in the graph is $Q = q$, in this case, we can treat such a graph as a graph with a single defect and the GSC has the form

$$C\left(G_{q-1}\right) = 2\frac{q\left(q-1\right)-2}{q^2\left(q-1\right)^2}. \qquad (17)$$

The greater the number of the nodes in the graph $G_{q-1}(Q)$ is, the less its GSC is affected by the defect: $C(G_{q-1}(q \gg 1)) \sim 2/q^2$, i.e. the GSC of a complete graph without one node tends to the GSC value of a complete graph, i.e. to zero, in other words, a small defect in the graph does not significantly change its complexity.

## 3.2 Vertex-transitive graphs

**Infinite square lattice.**

An infinite square lattice $G_{sq-\inf}$ is a vertex-transitive graph, i.e. a graph with the nodes having the same degree and the same position concerning the other nodes. The GSC of such a lattice is equal to the NodeSC of any of the graph nodes.

In Fig. 4a, on a part of an infinite square lattice, the edges and nodes of the first three steps of the GSC algorithm are shown, in Fig. 4b - the boundary region of a finite square lattice $G_{sq}$

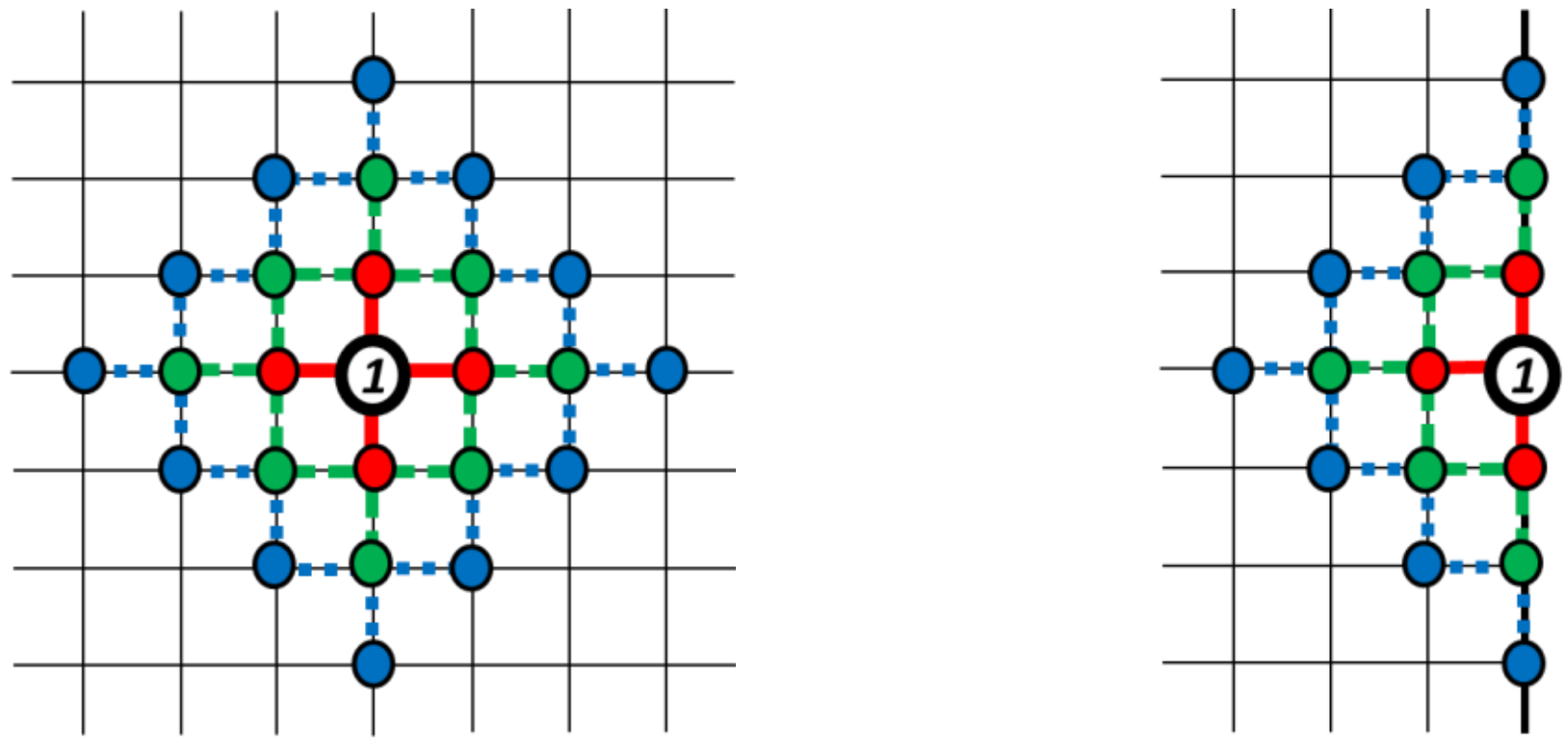

A b

Fig.4. Three steps of the GSC algorithm: a – for an arbitrary node of a vertex-transitive infinite square lattice $G_{sq-\inf}$, b – for a boundary node of a finite-size square lattice $G_{sq}$.

Shown in Fig. 4a, three steps of the algorithm allow us not only to find the first terms of the NodeSC sum for the infinite square lattice, but also make it possible to find the quantities $b_k$ and $s_k$ for each subsequent step of the algorithm in the following form

$$b_k = 4k^2\,, \quad s_k = 1 + 2k\left(k+1\right), \quad k = 1, 2, \ldots . \tag{18}$$

Thus, the NodeSC for node #1 is $C^1(G_{sq-\inf})$ as well as for any other node and, accordingly, the GSComplexity for the whole lattice is

$$C(G_{sq-\inf}) = C^1 = \sum_{k=1}^{\infty} \delta_k = \sum_{k=1}^{\infty} \frac{8k^2}{2k\left(k+1\right)\left[1+2k\left(k+1\right)\right]}\left[1 - \frac{8k^2}{2k\left(k+1\right)\left[1+2k\left(k+1\right)\right]}\right] = 0.931\,. \tag{19}$$

It should be noted, that when calculating the GSC, the passage to the limit of an arbitrarily large but finite size lattice to an infinite one is not trivial. Indeed, a large but finite-size lattice has peripheral nodes (in a layer adjacent to the graph boundary), for example, as in Fig. 4b. Their NodeSC (see Appendix A) is usually higher than the NodeSC for the internal nodes. In the finite size lattice $G_{sq}$, as the graph size increases, the fraction of the boundary nodes decreases and their contribution to the GSC also decreases and finally approaches zero. However, neglecting the NodeSC of the peripheral nodes is not always acceptable. For example, for a Bethe lattice (see Appendix B), for any finite size, the fraction of the peripheral nodes is about half of the total number of the nodes in the Bethe lattice and their contribution to the GSComplexity cannot be ignored.

**Prism graph.**

Let us now consider an example of finite vertex-transitive graphs with cycles – prism graphs [12], see Fig.5. In mathematical notation they are denoted as $Y_3 = GP\left(3,1\right)$, $Y_4 = GP\left(4,1\right)$, $Y_5 = GP\left(5,1\right)$,... $Y_p = GP\left(p,1\right)$, where $p = 3, 4, 5, \ldots$. As for the infinite square lattice $G_{sq-\inf}$, it is enough to find the NodeSC of any node to find the GSC of the graph.

Fig.5 shows prism graphs, as they are usually drawn, showing why they are called that way. The first three steps of the algorithm for each of the presented graphs with the beginning at node #1 are shown schematically.

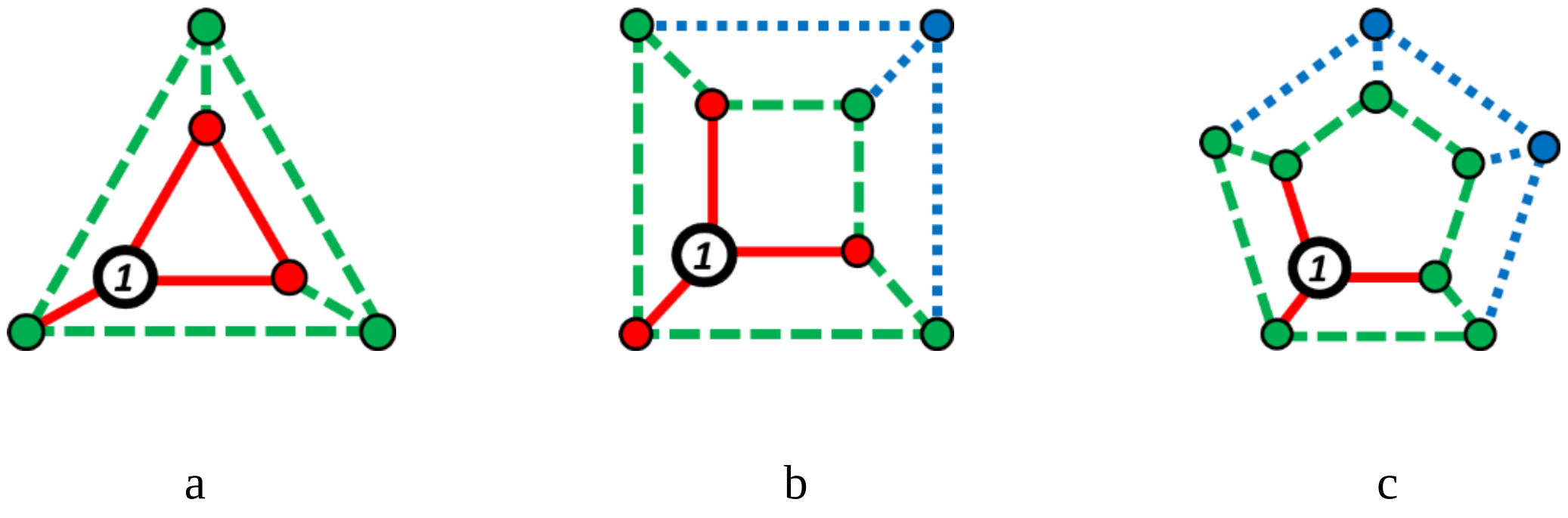

Fig.5. Prism graphs: a - $Y_3$, b - $Y_4$ and c-$Y_5$ with marked edges for three steps of the GSC algorithm.

As before for the infinite square lattice, it is possible to create relations for $b_k$ and $s_k$ for the prism graphs with $2p$ nodes, however, these relations will be different for even and odd $p$. The number of steps of the GSC algorithm for any size prism graph is $n = ceil((p+1)/2)$ (round up) steps. Expressions for the number of nodes and edges at each step of the GSC algorithm for even (20) and odd (21) graph prisms:

$$\begin{array}{lll} s_k = 4k, & b_k = 3+4k, & k = 1..n-1 \\ s_k = 4(n-1)+2, & b_k = 3+6(n-1)+3, & k = n \end{array}, \tag{20}$$

$$\begin{array}{lll} s_k = 4k, & b_k = 3+4k, & k = 1..n-2 \\ s_k = 4(n-2)+3, & b_k = 3+6(n-2)+7, & k = n-1 \\ s_k = 4(n-2)+4, & b_k = 3+6(n-2)+12, & k = n \end{array}. \tag{21}$$

The indicated relations are valid for graphs starting with $p > 6$, for smaller values, GSC can be calculated directly $C(Y_3) = 104/225 \approx 0.462$, $C(Y_4) = 0.74$, $C(Y_5) = 0.687$, $C(Y_6) = 0.865$.

Fig. 6a shows the GSComplexity of the prism graphs for sizes up to $p < 50$. When the size of the graph prisms grows, the GSCs of even and odd sizes get closer (see the insert of Fig.6a).

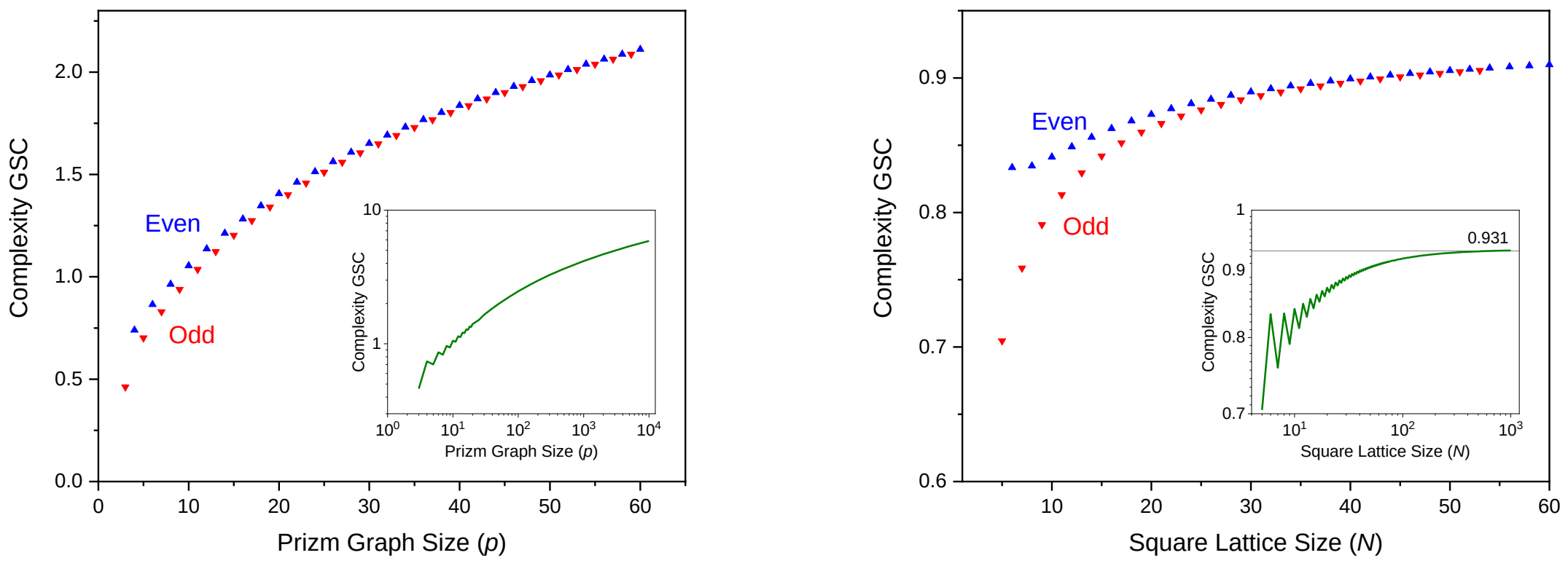

a b

Fig.6. The GSC of vertex-transitive graphs: a – prism graphs of various sizes and b – square periodical lattice (square lattice on a torus).

It is interesting to note that similar behavior is demonstrated by a square lattice of finite size $G_{sq}$ with periodical boundaries (square lattice on a torus), where for the small sizes, the noticeable GSC difference exists for even and odd sizes disappearing as the graph grows (see Appendix A). The results for such a graph are shown in Fig. 6b.

## 4. ARTIFICIAL RANDOM GRAPHS

The Erdős–Rényi graph $G(n, p)$ [13, 14] was modeled (binomial random graph model) using a given number of nodes $n$ and edge probability $p$, in such a graph [15], if $n \cdot p \to c > 1$, where $c$ is a constant, a graph $G(n, p)$ will almost certainly have a unique giant component containing a positive fraction of the nodes. In other words, when $p > 1/n$ a new object appears in the graph – a giant component.

The GSC calculation was performed for $G(n, p)$ with the number of nodes $n = 10^4$ for the concentrations near $p \simeq 1/n = 10^{-4}$, the results are shown in Fig. 7. Each point was averaged over 100 random realizations of the graph; the relative error was no more $0.09$ for concentrations $p < 8 \cdot 10^{-5}$ and $0.03$ for larger ones.

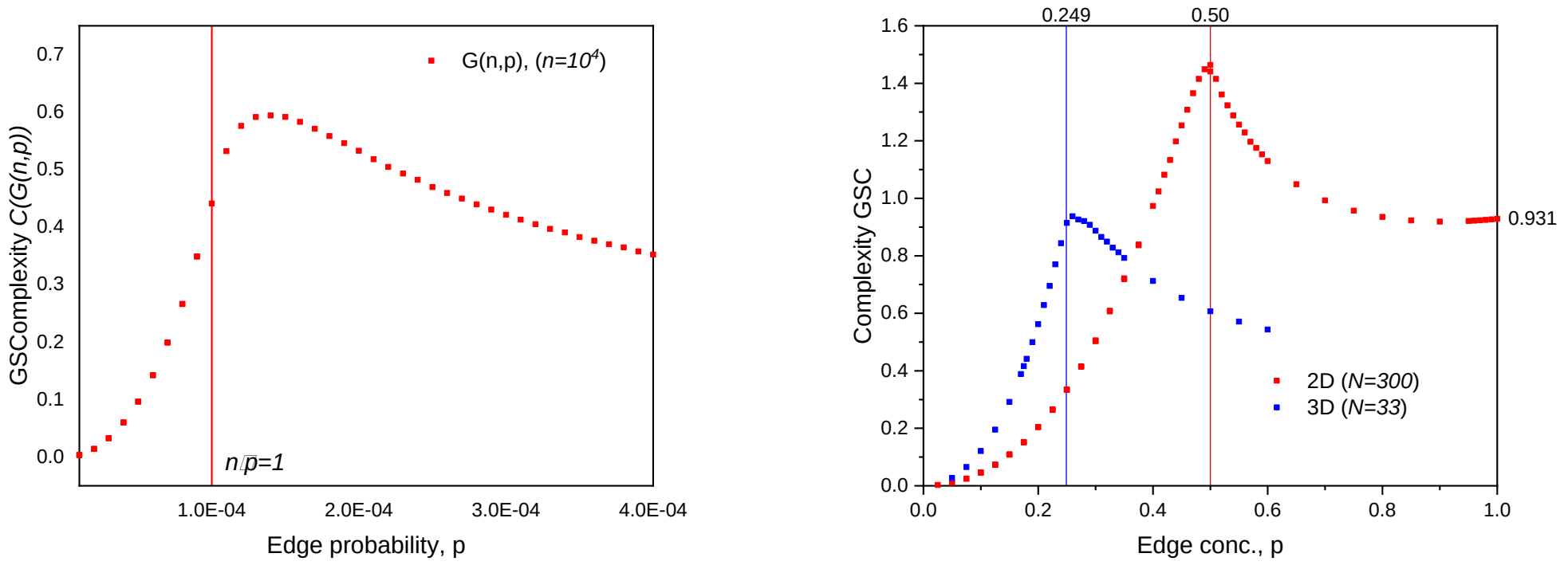

a b

Fig. 7. The GSComplexity of the systems with critical points: a – the Erdős–Rényi random graph $G(n, p)$ for the concentrations near to $p \simeq 1/n$, b – the percolation systems near the percolation threshold 2D square lattice and 3D cubic lattice.

Fig.7a clearly shows the GSC maximum of $C(G(n, p))$, which turned out to be in the region $p > 1/n = 10^{-4}$, $p \approx 1.3 \cdot 10^{-4}$. The observed peak definitely reflects the appearance of a giant component structure in the Erdős–Rényi graph [14].

In percolation structures [16], for example, in random graphs built on square or cubic lattices, the so-called infinite cluster appears, an analog of the giant component of the Erdos-Rényi random graph.

Consider a random graph $G_{2D}$ on a square finite-size lattice of size $N \times N$, the lattice has periodic boundary conditions, and such a lattice can be thought of as a square lattice on the torus surface.

Let consider a so-called bond problem, when each edge of the finite size lattice belongs to $G_{2D}$ with probability $p$ and, accordingly, does not belong to it with probability $1-p$. With an increase of probability $p$ at a certain value $p = p_c$, a percolation transition occurs, i.e. the infinite cluster appears. For a 2D square lattice $p_c = 0.5$, for a 3D cubic lattice $p_c \approx 0.249$ [16]. According to the general concept, the most complex structure of such a lattice graph occurs at the percolation threshold, and the infinite cluster itself is fractal.

The GSComplexity near and at the percolation threshold for 2D and 3D lattices are shown in Fig.7b. In the two-dimensional case, a clearly defined maximum is observed at the concentration of edges $p_c = 0.5$. For the cubic lattice, the GSC maximum also coincides with the value of the 3D percolation threshold – $p_c \approx 0.249$. The size of the lattice $N$ is indicated in Fig. 7b, the results were obtained by averaging 20 random realizations for each concentration in 2D and 3D, and the relative error was no more than 0.05.

When the probability $p$ tends to unity, the lattice transforms into a regular finite-size square (or cubic) lattice, while all its nodes are in the same conditions (because of periodic boundary conditions) and they have the same NodeSC. In Appendix A the NodeSC and the GSC of the 2D finite size lattice are found $C(G_{2D}) = 0.931$, presented results show the GSC value at $p \to 1$ has the same value (see Fig. 7b).

## 5. RESULTS FOR CARDIAC RHYTHM DATA

The idea of investigating the time series by mapping them to the complex networks (graphs) is very attractive. The Natural Visibility Graph (NVG) algorithm was proposed in [11]. A short description of NVG follows. Let $\{x(t_i), i=1..N\}$ be a time series of $N$ data, $t_i$ are in natural temporal ordering. The NVG [11] is created by mapping of a time series of $N$ data to a network (graph) of the $N$ nodes. The edge $(i, j)$ belongs NVG if on the time series plot $x(t_k)$ for all $t_k$ between $t_i$ and $t_j$ are below the line connecting $x(t_i)$ and $x(t_j)$, see Fig.8. Further extension of the NVG algorithm proposed in [10] is the Parametric Natural Visibility Graph (PNVG) algorithm, where mapping criteria depends on the introduced parameter $\alpha$ called "View angle". If in the natural temporal direction, the angle of the NVG edge to the vertical direction is wider than given parameter $\alpha$ the edge will belong to PNVG. Fig.8. has such edges marked by thick red (online) lines.

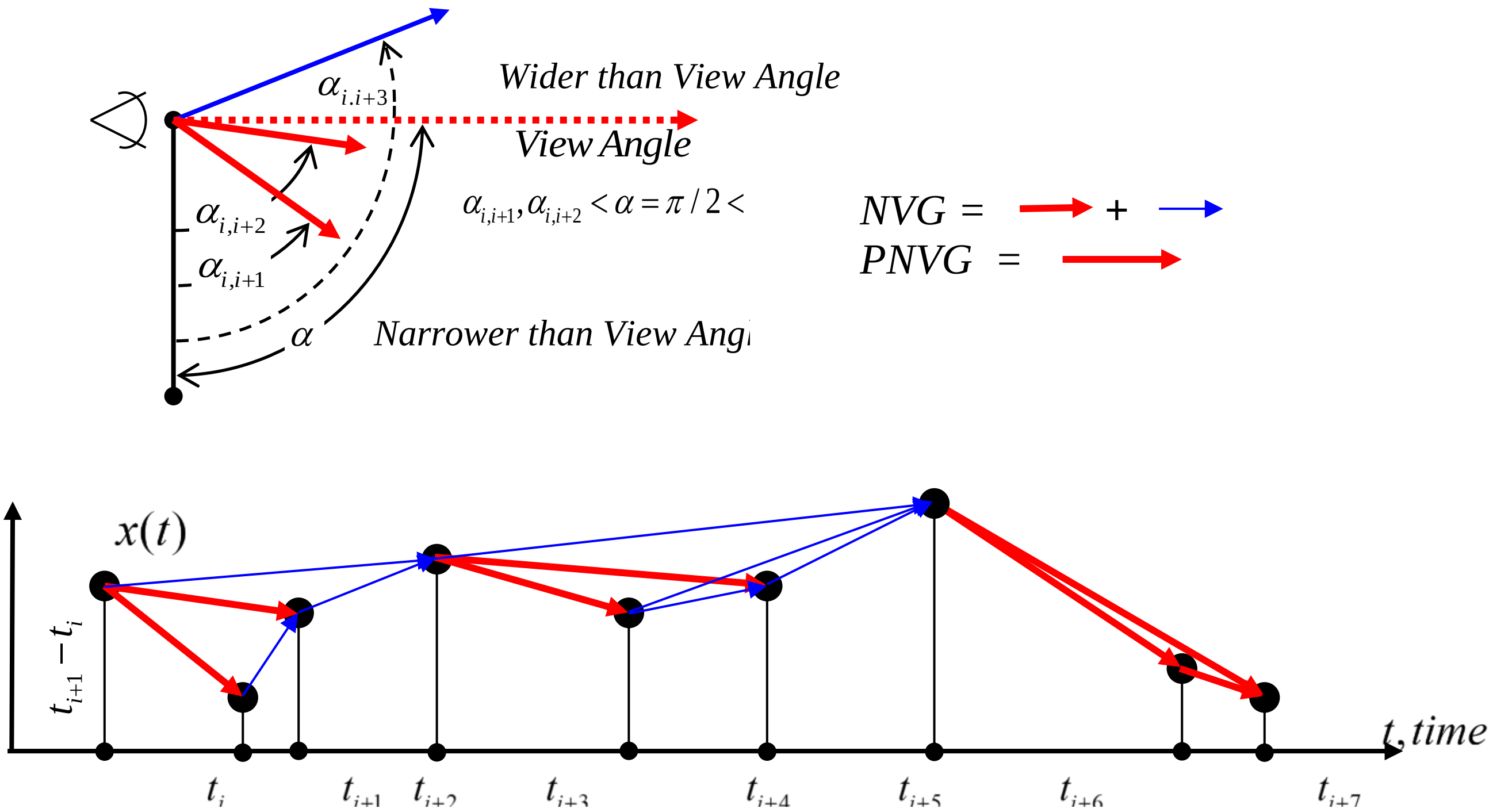

Fig. 8. Illustration of the Visibility Graph algorithms. Time series plot for NVG and PNVG algorithms. On the time axis events $t_i..t_{i+8}$ are marked. Upper left – the PNVG edge selection criterion for view angle $\alpha = \pi / 2$ applied to node $x(t_i)$. Lines $t_i, t_{i+1}$ and $t_i, t_{i+2}$ with $\alpha_{i,i+1}$ and $\alpha_{i,i+2}$ (thick lines, red online) belong to the PNVG( $\pi / 2$ ), the edge with $\alpha_{i,i+3}$ (thin lines, blue online) does not belong to the PNVG( $\pi / 2$ ). The NVG consists of thick (red online) and thin (blue online) links. The PNVG( $\pi / 2$ ) consists only of thick (red online) lines.

Next, we will give a short example of the calculation of the GSC of the complex network produced by $PNVG(\alpha)$ the algorithm for certain time series i.e $C(PNVG(\alpha))$. The results for different time series are presented in Fig. 9. To allow to the reader the convenience of the time series comparison we correct every GSC data set in the way $\bar{C}(PNVG(\alpha)) = C(PNVG(\alpha)) / C(PNVG(\pi))$ making the $\bar{C}(PNVG(\pi)) = 1$, we call it the relative GSC.

Just as a kind of reference, we use the artificial time series generated by a uniform random generator in the range [0..1] with 10000 data points. The shown in Fig. 9 the relative GSC results for such series are averaged over 100 realizations, the relative error at each $\alpha$ value was less than 0.04.

In the Fig.9, we present $\bar{C}(PNVG(\alpha))$ results for the experimental data – the series of the human heartbeat RR intervals. Data was taken from the PhysioNet [17] database. 54 series for healthy people (dataset nsr2db), 25 series for patients with Congestive Heart Failure (chr2db), and 83 series with Atrial Fibrillation (ltafdb) were examined. Each series has a different length and contains $6 \div 12 \times 10^4$ RR intervals.

The ECG data were processed by PhysioToolkit (*WFDB*) software, provided by PhysioNet [39]. Initially (when necessary) we annotate data using *wqrs* utility, then to extract a series of RR intervals we use *ann2rr* utility. To remove trend from RR intervals time series we use a detrending algorithm based on smoothness priors approach [18, 19]. The regularization parameter was set to $\lambda_{reg} = 15$. Finally, we correct RR intervals time series so that their mean value is equal to 1.

For each time series above, for view angles $\alpha$ in the range $[1.5, 1.7]$ i.e. near to horizontal view angle we construct PNVG ( $\alpha$ ) and further compute $\bar{C}(PNVG(\alpha))$. Then we calculate average values over each dataset type $\left\langle \bar{C}(PNVG(\alpha)) \right\rangle$.

Each type of cardiac rhythm produces its particular $\bar{C}(PNVG(\alpha))$, in Fig.9 we present averaged rhythm data $\left\langle \bar{C}(PNVG(\alpha)) \right\rangle$ for every type of considered cardiac rhythm. The shapes of the curves in Fig.9 are visually distinguishable one from another. The maximum of averaged relative GSC $\left\langle \bar{C}(PNVG(\alpha)) \right\rangle$ upon $\alpha$ has the maximum near to $\alpha \approx \pi / 2$ for all processed time series. It is worth mentioning here that in the parametric visibility graph, the second-order phase transition was observed [20] at $\alpha \approx \pi / 2$, and here near this point the GSComplexity manifests the maximum as was demonstrated for the systems with the phase transitions in the previous chapter.

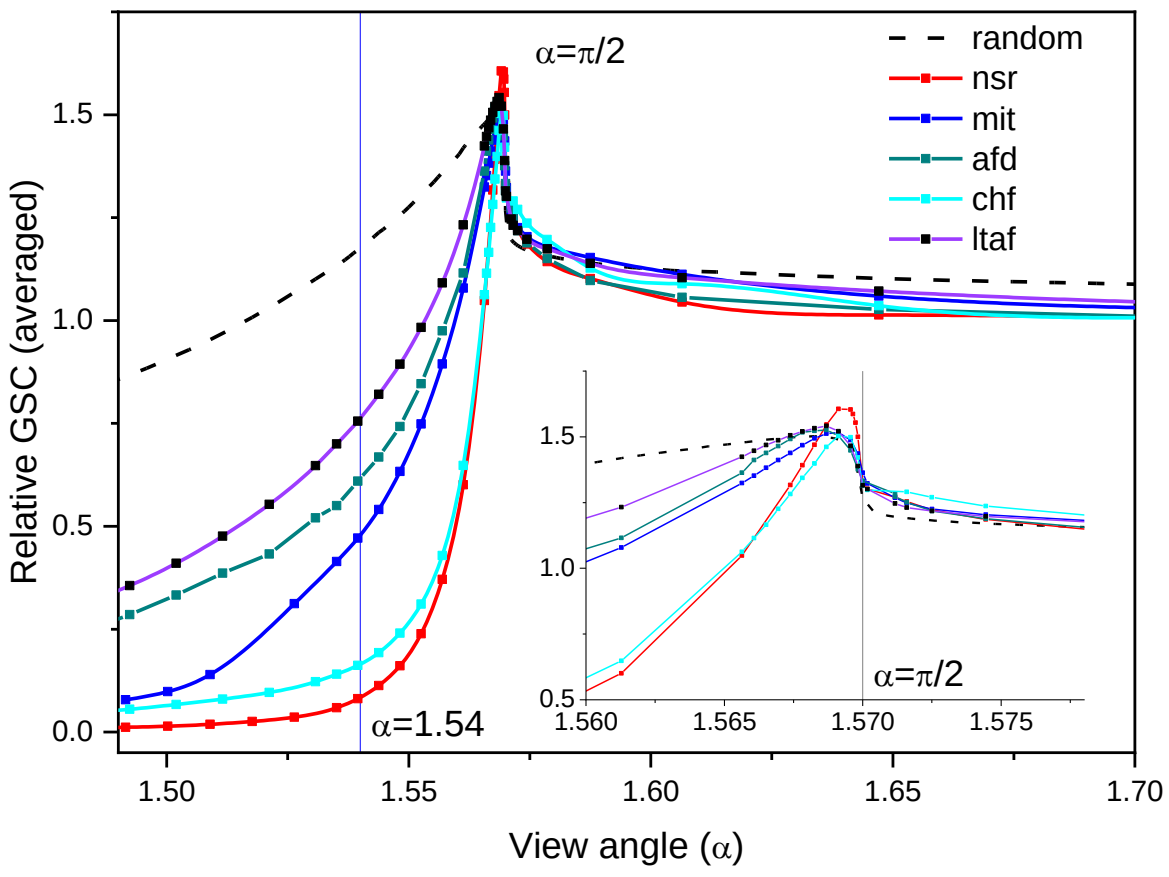

Fig. 9. The averaged relative GSC $\langle \bar{C}(PNVG(\alpha)) \rangle$ as a function of the view angle $\alpha$ for the different types of the time series (shown in the plot legend).

Unfortunately, each RR intervals time series produces its own $\bar{C}(PNVG(\alpha))$ shape that is often different from the average shape shown in Fig.9, but there is always the peak at $\alpha \approx \pi / 2$. Also from Fig.9 one can see that the shapes of the relative GSC for different sets and the GSC for uniform random data set are almost the same for $\alpha > \pi / 2$, but at $\alpha \leq \pi / 2$ curves are simple to distinguish. For instance, at $\alpha \approx 1.54$ the averaged values $\bar{C}(PNVG(\alpha))$ are different for every RR intervals set and significantly differ from uniform random time series.

The insert in Fig.9 shows in detail the data near $\alpha \approx \pi / 2$. For nsr and chf series the GSC maximum is a bit right relative to mit, afd and ltaf series but the slope at $\alpha \approx 1.565$ is clearly different, i.e. the nsr and chf time series are rising twice as the mit, afd and ltaf series do.

Therefore, the GSC of the PNVG for cardiac rhythm data could be one more possible parameter that allows to the doctors to distinguish at least partially different heart diseases.

## 6. CONCLUSIONS AND DISCUSSIONS

Here we have introduced a quantitative measure of the graph (complex network, etc.) structural complexity based on a procedure similar to the renormalization process, considering the difference between actual and averaged graph structures on different scales. We applied this approach to find the structural complexities for various graph types – the deterministic infinite and finite size graphs, artificial graphs of different natures including percolation structures, and the time series of cardiac rhythms mapped to complex networks using the parametric visibility graph algorithm.

The proposed concept of the graph structural complexity corresponds to qualitative comprehension of the complexity. The more complex is the graph structure, the greater the complexity of the graph is. The structural complexity of the empty and complete graphs (having no structure) is zero, the complexity reaches a maximum near the formation of a giant component in the graph or at the percolation threshold for 2D and 3D square lattices when a giant cluster having a fractal structure emerges. The latter reveals that the graph structural complexity allows us to detect and study the processes similar to second-order phase transitions in complex networks.

The proposed GSC algorithm can handle weighted graphs also. Even it can handle the graphs with the weights assigned to the nodes. To do this, a transition to be made, one of the possible ones is to assign to the edges the weight equal to the sum of the partial weights of the connected nodes divided by the degree of such node. This makes it possible to find, for instance, the complexity of the image as the GSC of the regular (two-dimensional or multidimensional) lattice with weighted nodes, or even the complexity of the unregular weighted nodes network.

It is possible as a part of the GSC algorithm to introduce a new node centrality Index, characterizing the structural complexity of a certain node within the graph structure, i.e. to find the node complexity distribution over the graph. This centrality index can serve, for example, as a good auxiliary or generalization to the local clustering coefficient. The structural complexity of individual nodes provides another new ranking manner for the graph nodes. We believe that the changes in the structural complexity of complex social networks over time can give the evidence or be the indicator to forthcoming often hidden social processes. Possible applications also belong to the context of machine learning-based algorithms or artificial intelligence that can utilize calculated complexities.

Being an easily computable measure, it might help to reveal different features of complex systems and processes. Undoubtedly, the following studies are necessary to demonstrate the usability of the proposed graph structural complexity to many different phenomena of the real world.

## 7 ACKNOWLEDGEMENT

The author would like to thank Dr. Igor Bezsudnov for the numerous fruitful discussions and assistance in calculations and Prof. D. Lande for discussions on the issues raised.

# APPENDIX A

## Finite Size Square Lattice

It is possible to construct several types of square lattice: Type 1 – finite-size lattice with periodic boundary conditions (discussed in detail in Chapter 3), Type 2 – finite-size lattice produced by successive steps of the GSC algorithm (see Fig. 4a), Type 3 – lattice finite size square shape without periodic boundary conditions.

The type 1 square lattices are vertex-transitive (see Chapter 2), and the NodeSC of any node is equal to the GSComplexity of the entire lattice. Fig. A1 (corresponds to Fig., 6b) shows the numerical simulation of the GSC of the type 1 lattices.

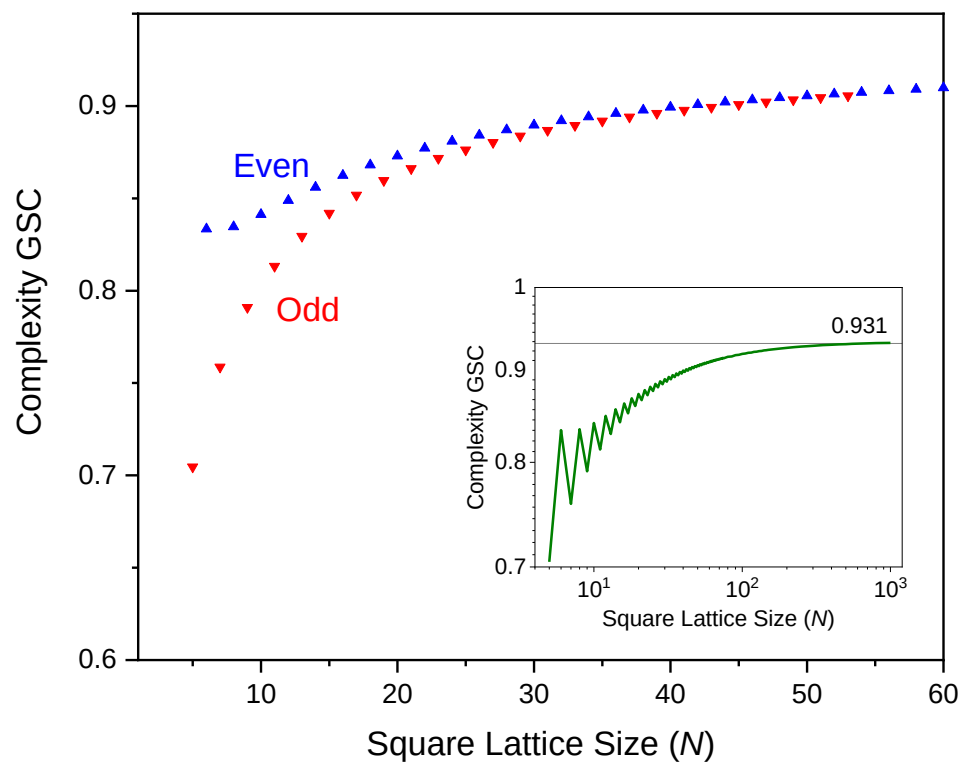

Fig. A1. The GSC for the finite size square lattice of size $N \times N$ with periodic boundary conditions for sizes $N < 100$, the inset for $N < 1000$.

For small sizes of Type 1 lattices the GSC is bouncing between even and odd sizes, and for large sizes $N$ (inset Fig. A1) the Type 1 GSC tends to the GSC value of an infinite square lattice - 0.931.

In a finite-size square lattice of the Type 2 or Type 3 i.e. without periodical boundary conditions, one can distinguish "internal" nodes – their quantity $\sim N^2$ and boundary ones $\sim N$. The numerical value of the NodeSC of internal nodes for a large lattice size turns out to be close to 0.931. The first steps of the GSC algorithm for computing the NodeSC of a Type 3 lattice boundary node are shown in Fig. 4b.

The number of nodes and edges of the first steps are $s_1 = 4, \quad b_1 = 3, \; s_2 = 9, \quad b_2 = 10$, $s_3 = 16, \quad b_3 = 21$. For the $k$-th step $p_k$ is

$$s_k = k^2, \quad b_k = k(2k+1), \quad k = 1, 2, \ldots, \tag{A.1}$$

$$p_k = \frac{b_k}{m_k} = 2\frac{k(2k+1)}{k^2(k^2-1)}, \tag{A.2}$$

and, to sum up,

$$\delta_k = 2\frac{k(2k+1)}{k^2(k^2-1)}\left[1 - 2\frac{k(2k+1)}{k^2(k^2-1)}\right], \tag{A.3}$$

we find

$$C^1 = \sum_{k=1}^{\infty} \delta_k = 1.313. \tag{A.4}$$

Indeed, for a square lattice of finite size the NodeSC of the boundary node is 1.313 is greater than the internal NodeSC - 0.931. However, the contribution to the GSC of the boundary nodes at $N \to \infty$ tends to zero at large sizes, since the fraction of boundary nodes in the lattice under consideration is of the order $\sim 1/N$ of the total number of lattice nodes.

## APPENDIX B

### Bethe lattice.

The Bethe lattice $G_B(Z)$ is an infinite cycle-free graph with a constant coordinate number $Z$, Fig. B1a shows a fragment of such a graph for $Z = 3$.

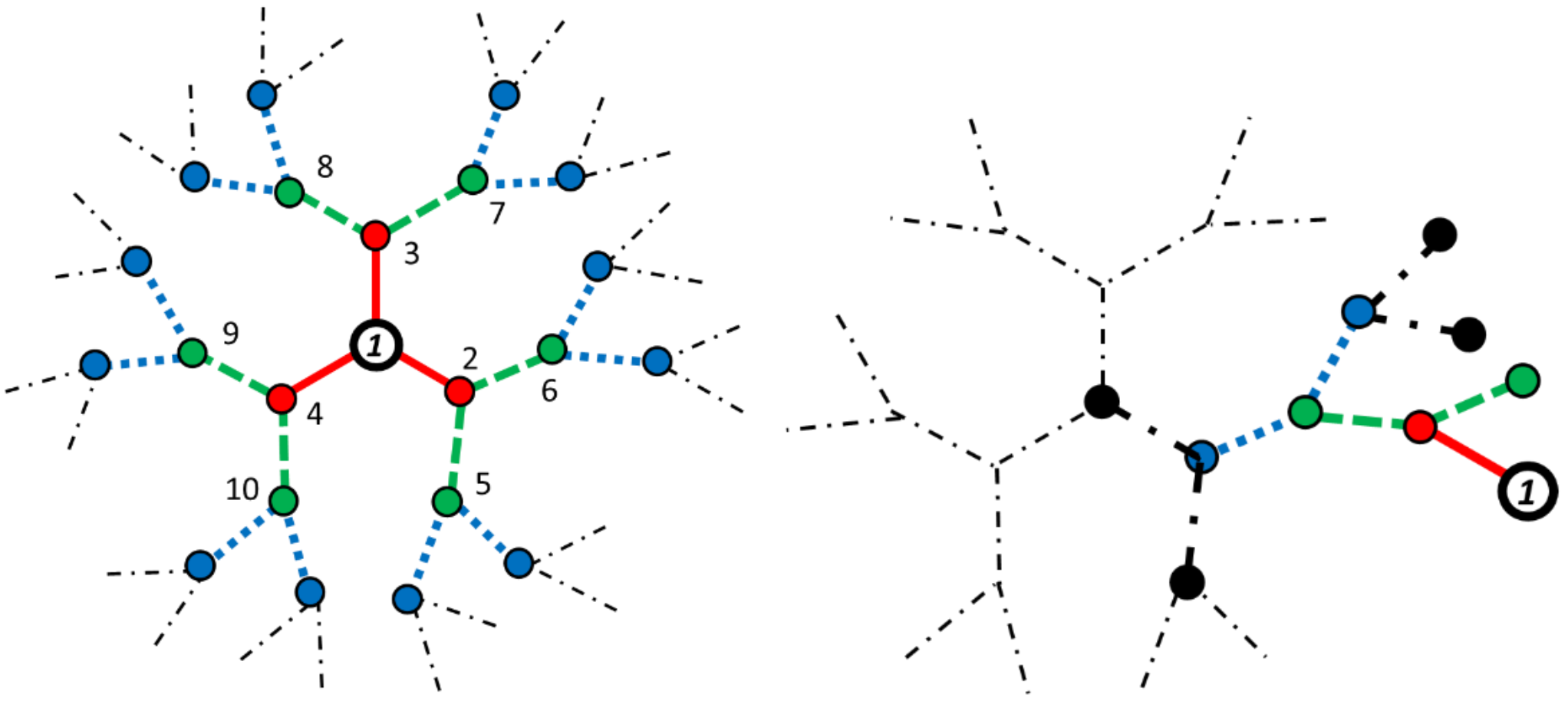

a b

Fig. B1. First steps of the GSC algorithm for the Bethe lattice $G_B(3)$: a – part of an infinite lattice starting from node #1, 3 steps, b – the boundary node of the finite Bethe lattice, 4 steps.

Let's find the NodeSC for node #1 (see Fig. B1a), below for convenience we omit the superscript in the notation. During the first step of the GSC algorithm for the Bethe lattice $k = 1$, in addition to node #1, three more are reachable: #2, #3, and #4 (see Fig. B1a). Thus, the first subgraph $g_1$ consists of four nodes $\sigma_1 = \{1,2,3,4\}$, $s_1 = 4$, and three edges $\beta_1 = \{1-2, 1-3, 1-4\}$, $b_1 = 3$, and the edges quantity in the complete graph $m_1 = s_1(s_1 - 1)/2 = 6$, the graph density is $p_1 = b_1 / m_1 = 1/2$, and the variance is $\delta_1 = p_1(1 - p_1) = 1/4$.

At the second step of the GSC algorithm, the set of nodes is $\sigma_2 = \{1, 2, ..., 10\}$, $s_2 = 10$,, and the set of edges is $\beta_2 = \{1-2, 1-3, 1-4, 2-6, 2-5, 3-7, 3-8, 4-9, 4-10\}$, $b_2 = 9$. Density and variance are equal to $p_2 = b_2 / m_2 = 1/5$ and $\delta_2 = p_2(1 - p_2) = 4/25$ respectively. For the $k$-th step $b_k(Z=3) = 3(2^k - 1)$, $s_k(Z=3) = b_k(Z=3) + 1$.

The general expression for the variance at each step of the GSC algorithm can also be found and the GSComplexity can be written in the form

$$C_{Bethe}(Z=3) = C^1 = \sum_{k=1}^{\infty} \delta_k = \sum_{k=1}^{\infty} \frac{b_k}{s_k(s_k-1)} \left(1 - \frac{b_k}{s_k(s_k-1)}\right) = 0.576 . \quad \text{(B.1)}$$

For an arbitrary coordination number $Z$

$$b_k(Z) = Z \frac{(Z-1)^k - 1}{Z-2}, \quad s_k(Z) = b_k(Z) + 1. \quad \text{(B.2)}$$

And thus, the GSC of Bethe lattices with an arbitrary coordination number $Z$ will be $C(G_B(4)) = 0.399$, $C(G_B(5)) = 0.268$, $C(G_B(6)) = 0.234$. $Z \gg 1$ The GSC can be approximated by a power function

$$C(G_B(Z)) \approx 2 \cdot Z^{-1}. \quad \text{(B.3)}$$

The GSC of the Bethe lattice was found for the case of an infinite lattice, i.e. with an infinite number of steps from the central node, for an infinite lattice the NodeSC's of all nodes are the same.

In this case, when dealing with the finite size Bethe lattice $G_B(Z,M)$, i.e. the graph constructed by $M$ branching steps originated by the central node with the coordinate number $Z$, it would be incorrect to suppose that the GSC of $G_B(Z,M)$ is equal or near to the GSC of infinite Bethe lattice.

Figure B1b shows the first steps of the GSC algorithm for a boundary node of a finite-size Bethe lattice. For the first step, the resulting subgraph is complete $\delta_1 = 0$. For the second step, the number of nodes and edges – $s_2 = 4, \quad b_2 = 3$ and thus $\delta_2 = 1/4$. For the third - $s_3 = 6, \quad b_3 = 5$ and $\delta_3 = 2/9$. For the fourth step $s_4 = 10, \quad b_4 = 9$ and $\delta_4 = 4/25$.. Thus, already at the fourth step of the NodeSC computation for the boundary node of the Bethe lattice is higher than the NodeSC for the infinite Bethe lattice

$$\sum_{k=1}^{4} \delta_k = 0 + \frac{1}{4} + \frac{1}{3} + \frac{4}{25} \approx 0.61 > C\left(G_B(3)\right). \tag{B.4}$$

The quantity of the boundary nodes relative to internal ones for a finite size Bethe lattice tends to have finite value

$$\frac{s_{k+1} - s_k}{s_k} \to Z - 2, \quad k \to \infty, \tag{B.5}$$

Therefore, (B.4) indicates that the share of the boundary nodes is proportional to the total number of the nodes, and their contribution to the GSComplexity is significant for an arbitrarily large but finite-size Bethe lattice. Indeed, as numerical simulation shows, with the large Bethe lattice sizes the GSC becomes equal to $C\left(G_B(3,M)\right)\approx 1.04$, which is greater than that for the infinite case $C\left(G_B(3)\right)=0.576$.

Fig. B2a shows $C^1\left(G_B(3,M)\right)=\sum_{k=1}^{M}\delta_k$, i.e. the NodeSC of the central node of the finite size Bethe lattice vs. the branching steps $M$ of the built graph.

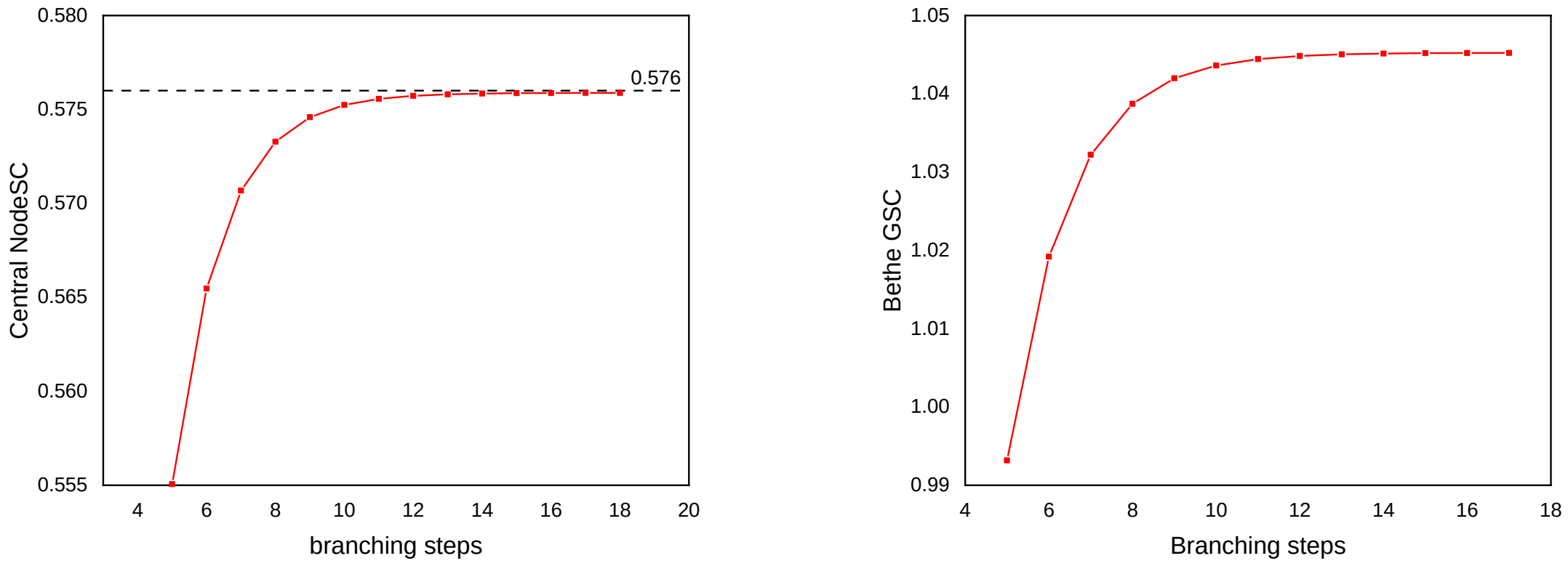

a b

Fig. B2. The GSComplexities for different branching steps $M$ of the Bethe lattice $G_B(3,M)$: a – the central NodeSC $C^1\left(G_B(3,M)\right)$, b – the GSC $G_B(3,M)$ of the finite size Bethe lattice.

As one can see, even the NodeSC of the central node (and neighbor nodes) quickly rises and becomes equal to $C\left(G_B(3)\right)=0.576$, the GSC of the whole Bethe lattice (see Fig. B2b) tends to $C\left(G_B(3,M)\right)\approx 1.04$, i.e. presenting to us the example of the incorrect passage of the GSC of finite size graph to the infinite graph having the same structure.

## REFERENCES

1. Albert, R. and A.-L. Barabási, *Statistical mechanics of complex networks.* Reviews of Modern Physics, 2002. **74**(1): p. 47-97.
2. Estrada, E., *The structure of complex networks: theory and applications*. 2012, USA: Oxford University Press.
3. Jensen, H.J., *Complexity science: the study of emergence*. 2022, USA: Cambridge University Press.
4. Lloyd, S., *Measures of complexity: a nonexhaustive list.* IEEE Control Systems, 2001. **21**(4): p. 7-8.
5. Lau, Z.J., et al., *Brain entropy, fractal dimensions and predictability: A review of complexity measures for EEG in healthy and neuropsychiatric populations.* European Journal of Neuroscience, 2022. **56**(7): p. 5047-5069.
6. Bagrov, A.A., et al., *Multiscale structural complexity of natural patterns.* Proceedings of the National Academy of Sciences, 2020. **117**(48): p. 30241-30251.
7. Peng, C.K., et al., *Mosaic organization of DNA nucleotides.* Physical Review E, 1994. **49**(2): p. 1685-1689.
8. Peng, C.K., et al., *Quantification of scaling exponents and crossover phenomena in nonstationary heartbeat time series.* Chaos: An Interdisciplinary Journal of Nonlinear Science, 1995. **5**(1): p. 82-87.
9. Skiena, S.S., *Sorting and Searching*, in *The Algorithm Design Manual*. 2012. p. 103-144.
10. Bezsudnov, I.V. and A.A. Snarskii, *From the time series to the complex networks: The parametric natural visibility graph.* Physica A: Statistical Mechanics and its Applications, 2014. **414**: p. 53-60.
11. Lacasa, L., et al., *From time series to complex networks: The visibility graph.* Proceedings of the National Academy of Sciences, 2008. **105**(13): p. 4972-4975.
12. Read, R.C., ; Wilson, R. J., *An Atlas of Graphs*. 2004, Oxford, England: Oxford University Press.
13. Dorogovtsev, S.N., A.V. Goltsev, and J.F.F. Mendes, *Critical phenomena in complex networks.* Reviews of Modern Physics, 2008. **80**(4): p. 1275-1335.
14. Erdös, P. and A. Rényi, *On Random Graphs I.* Publicationes Mathematicae Debrecen, 1959.
15. Erdos, P.L. and A. Rényi, *On the evolution of random graphs.* Transactions of the American Mathematical Society, 1984. **286**: p. 257-257.
16. Stauffer, D. and A. Aharony, *Introduction To Percolation Theory: Second Edition (2nd ed.)*. 1992, London: Taylor & Francis.

17. Goldberger, A.L., et al., *PhysioBank, PhysioToolkit, and PhysioNet.* Circulation, 2000. **101**(23).
18. Gersch, W., *Smoothness Priors*, in *New Directions in Time Series Analysis, , Part II* 1991, Springer-Verlag press.
19. Tarvainen, M.P., P.O. Ranta-aho, and P.A. Karjalainen, *An advanced detrending method with application to HRV analysis.* IEEE Transactions on Biomedical Engineering, 2002. **49**(2): p. 172-175.
20. Snarskii, A.A. and I.V. Bezsudnov, *Phase transition in the parametric natural visibility graph.* Physical Review E, 2016. **94**(4).